\documentclass[lettersize,journal]{IEEEtran}
\usepackage{amsmath,amsfonts}
\usepackage{algorithmic}
\usepackage{algorithm}
\usepackage{array}
\usepackage[caption=false,font=normalsize,labelfont=sf,textfont=sf]{subfig}
\usepackage{textcomp}
\usepackage{stfloats}
\usepackage{url}
\usepackage{verbatim}
\usepackage{graphicx}
\usepackage{cite}
\usepackage[utf8]{inputenc}
\usepackage{amssymb}
\usepackage{tikz}
\usepackage{eso-pic}
\usepackage[colorlinks,allcolors=black,urlcolor=blue]{hyperref}

\AddToShipoutPicture{%
  \sffamily\scriptsize
  \setlength\unitlength{1mm}%
  \AtPageUpperLeft{%
    \put(40,-3){\parbox[t]{0.8\textwidth}{%
        \centering
        This article has been accepted for publication in
        IEEE Photonics Technology Letters.  This is the author's
        version which has not been fully edited and content may
        change prior to final publication.
        Citation information:
        \href{https://dx.doi.org/10.1109/LPT.2022.3195433}{%
          DOI 10.1109/LPT.2022.3195433}
    }}%
  }%
  \AtPageLowerLeft{%
    \put(40,3){\parbox[b]{0.8\textwidth}{%
        \tiny
        \centering
        © 2022 IEEE. Personal use is permitted, but
        republication/redistribution requires IEEE permission.
        See \url{https://www.ieee.org/publications/rights/index.html}
        for more information.
    }}%
  }%
}

\begin{document}

\title{Potential impact of CV-QKD integration on classical WDM network capacity}

\author{%
  Cédric Ware,~\IEEEmembership{Senior Member,~IEEE,}
  Raphaël Aymeric,
  Chaima Zidi,
  Mounia Lourdiane
  \thanks{%
    Cédric Ware and Raphaël Aymeric are with LTCI, Télécom Paris,
    Institut Polytechnique de Paris, Palaiseau, France.
    Mounia Lourdiane is with SAMOVAR, Télécom SudParis,
    Institut Polytechnique de Paris, Évry, France.
    Chaima Zidi was formerly with both institutions mentioned above.
  }%
}

\markboth{IEEE Photonics Technology Letters, accepted.}%
         {Ware \MakeLowercase{\textit{et al.}}:
           Indications of impact on network capacity of CV-QKD coexistence
           with classical traffic}

\IEEEpubid{\parbox[t]{0.9\textwidth}{© 2022 IEEE.  Personal use of this material is permitted.  Permission from IEEE must be obtained for all other uses, in any current or future media, including reprinting/republishing this material for advertising or promotional purposes, creating new collective works, for resale or redistribution to servers or lists, or reuse of any copyrighted component of this work in other works.}}

\maketitle

\begin{abstract}
  Continuous-variable quantum key distribution (CV-QKD) could
  allow QKD and classical optical signals physically sharing
  the same optical fibers in existing networks.  However, Raman
  scattering imposes a limit on the optical power, which in turn
  impacts the network capacity for classical traffic in presence of CV-QKD.
  Network-planning simulations indicate that maxing out the CV-QKD capacity
  in an optical link can adversely impact its classical capacity.
  Although preliminary, these results show that designing a
  mixed classical and CV-QKD network will require dedicated planning
  heuristics and tools that specifically seek a compromise between
  classical and CV-QKD traffics.
\end{abstract}


\section{Introduction}
\IEEEPARstart{Q}{uantum-key distribution} (QKD)
is often presented as a future-proof physical-security solution
to the key distribution problem.  Within that field,
continuous-variable (CV) QKD uses the same type of
coherent-detection optical receivers as classical data traffic,
which gives it the potential not only to share hardware with
classical channels, but also---and of great interest to
network operators---to physically coexist with them in the same
optical fibers through wavelength-division multiplexing (WDM),
using the intrinsic spectral-filtering property of coherent detection%
~\cite{kumar-2015-coexistence,eriksson-2019-wdm-cvqkd}.
We recently demonstrated experimentally a symbiotic operation of
CV-QKD and classical communication where the latter is actually
used as a reference signal for phase and frequency recovery on
the quantum channel%
~\cite{aymeric-OFC-2022-symbiotic}.
The big drawback of CV-QKD compared to alternatives is its shorter range:
beyond a certain propagation length, due to attenuation in the fiber,
the key rate drops sharply;
In \cite{laudenbach-cvqkd-theory-2018} the theoretical key rate
is calculated depending on a number of parameters,
including attenuation, but also various noise-generating processes.
Among those, Raman scattering results in a crosstalk-like noise between
WDM channels, a significant limiting factor if CV-QKD is to
coexist with classical traffic.
The aforementioned references%
~\cite{kumar-2015-coexistence,eriksson-2019-wdm-cvqkd,aymeric-OFC-2022-symbiotic,laudenbach-cvqkd-theory-2018}
study CV-QKD key rate, including in the case
of coexistence with intense WDM channels, but for a single optical link,
not in a complete network.  On the other hand, quantum network capacity
and design tools have been studied~\cite{azuma-2021-quantum-network-design},
but without coexistence.

This paper searches a limit on the coexistence of both signals
within a network.
Since CV-QKD is impeded by the optical power of classical WDM channels
circulating alongside, conversely, enabling a certain amount of
CV-QKD traffic in a fiber requires setting a limit on optical power,
thus capping the number of WDM channels available in a given fiber for
classical traffic, which in turn impacts the overall network capacity.
Our simulations indicate that unless precautions are taken, above a certain
threshold of CV-QKD
offered traffic, optical links become unable to sustain classical
traffic, effectively dropping out of the network.  This sharply increases
the blocking probability on the classical traffic, defined as the ratio
of traffic blocked (that the network fails to carry) over the total
traffic requested.

Our methodology and choice of parameter values are described in
Sec.~\ref{sec/methodology}, yielding the simulation results
shown in Sec.~\ref{sec/results}.  We then discuss our interpretation,
limits of these preliminary results, and especially ways to further
investigate this problem in Sec.~\ref{sec/discussion},
before concluding the paper.

\section{Methodology and parameters}
\label{sec/methodology}
\IEEEpubidadjcol

\subsection{CV-QKD and classical signal parameters}

\begin{table}
  \caption{%
    Physical, technological and simulation parameters.%
    \label{tab/parameters}%
  }
  \centering
  \begin{tabular}{lcc}
    \hline
    Maximum classical WDM channels per fiber & $N$ & 40\\
    Classical data rate per WDM channel & $\mathcal{R}$ & 100~Gbit/s\\
    Classical optical power per WDM channel & $P_{\text{opt}}$ & 0~dBm\\
    Optical fiber attenuation & $\alpha_L$ & 0.2~dB/km\\
    Insertion loss on quantum channel & $\alpha_0$ & 2~dB\\
    CV-QKD symbol rate & $f_{\text{sym}}$ & 1~Gbaud\\
    CV-QKD reconciliation efficiency & $\beta$ & 0.95\\
    CV-QKD detection type & $\mu$ & 2 (heterodyne)\\
    CV-QKD receiver fixed noise & $\xi_0$ & $10^{-3}$~SNU\\
    CV-QKD Raman noise efficiency & $\xi_R$ & variable, SNU/W\\
    Maximum paths per node pair in RWA & $k$ & 5\\
    Distance scaling factor & $\Lambda$ & 0.01--0.2\\
    \hline
  \end{tabular}
\end{table}

We chose some ``reasonable'' values, summarized in
table~\ref{tab/parameters}, for a classical WDM network simulation,
as well as physical and technological parameters required to calculate
the CV-QKD key rate (considered in
the asymptotic regime of an infinite number of quantum state exchanges).
We assume that each optical link can carry CV-QKD traffic in a dedicated
spectral window separate from classical WDM channels, with a
secure-key rate of $K = f_{\text{sym}} \cdot r$%
~\cite[Eq.~(2.1)]{laudenbach-cvqkd-theory-2018},
where $f_{\text{sym}}$ is the quantum channel's symbol rate and $r$ the
secret fraction (bits per symbol transmitted),
whose optimal value can be calculated numerically as a function of
channel transmittivity $T$, excess noise $\xi$ (defined here at
channel output), efficiency of
information reconciliation $\beta$ and detection type $\mu$%
~\cite[Sec.~10]{laudenbach-cvqkd-theory-2018}, with a
100-\% efficient receiver.
For a link of length $L$ with $n$ WDM classical signals, we assume:
\begin{subequations}
  \begin{align}
    T &= 10^{-\alpha_{\text{dB}} / 10}\\
    \alpha_{\text{dB}} &= \alpha_0 + \alpha_L \cdot L\\
    \xi &= \xi_0 + \xi_R \cdot n \cdot P_{\text{opt}}
    \label{eq/quantum-noise}
  \end{align}
\end{subequations}
with $\alpha_0$ a fixed insertion loss and $\alpha_L$ the fiber's attenuation.
We use a simple linear model for $\xi$ as a function of the per-WDM-channel
optical power $P_{\text{opt}}$,
with $\xi_0 = 10^{-3}$ shot-noise units (SNU)
corresponding to the detection noise (untrusted),
a value optimistically chosen, as is $f_{\text{sym}} = 1$~Gbaud.
(In \cite{aymeric-OFC-2022-symbiotic}, we managed
$f_{\text{sym}} = 250$~Mbaud and $\xi_0 \simeq 9\cdot 10^{-2}$.)
One major uncertainty is the choice of the Raman efficiency $\xi_R$,
which will be specifically investigated
in Sec.~\ref{sec/raman-coefficient}.

\subsection{Network simulation, topology and traffic matrix}

\begin{figure}
  \centering
  \begin{tikzpicture}[scale=0.7,font=\scriptsize]
    \coordinate (Madrid) at (-3.6919444,40.9);
    \coordinate (Barcelona) at (2.1769444,41.3);
    \coordinate (Valencia) at (-0.375,40.2);
    \coordinate (Sevilla) at (-6.1,39.6);
    \coordinate (Zaragoza) at (-1,41.65);
    \coordinate (Malaga) at (-4.3,38.9);
    \coordinate (Murcia) at (-1.2,39.0);
    \draw (Barcelona)
    -- node [below,sloped] {$\Lambda\times 303$~km} (Valencia)
    -- node [right] {$\Lambda\times 177$~km} (Murcia)
    -- node [above,sloped,xshift=2mm] {$\Lambda\times 323$~km} (Malaga)
    -- node [below left,xshift=4mm] {$\Lambda\times 158$~km} (Sevilla)
    -- node [above,sloped] {$\Lambda\times 391$~km} (Madrid)
    -- node [above,sloped] {$\Lambda\times 272$~km} (Zaragoza)
    -- node [below,sloped,xshift=-2mm] {$\Lambda\times 257$~km} (Barcelona)
    (Madrid) -- node [above,sloped,xshift=1mm] {$\Lambda\times 302$~km} (Valencia);
    \foreach \id/\o/\r/\c/\t in {%
      1/above left/100/Madrid,3/above/70/Barcelona,%
      4/left/225/Valencia,7/left/135/Sevilla,%
      2/above/north/Zaragoza,%
      6/above/120/Malaga/Málaga,5/right/east/Murcia} {
      \node [draw,circle,fill=white,inner sep=1pt] (node \c) at (\c) {\id};
      \node [\o] at (node \c.\r) {``\t''};
    }
  \end{tikzpicture}
  \caption{Network topology with 7 nodes and
    8 bidirectional (16 unidirectional) links.
    Distances are uniformly scaled by factor $\Lambda$.%
    \label{fig/topology}}
\end{figure}
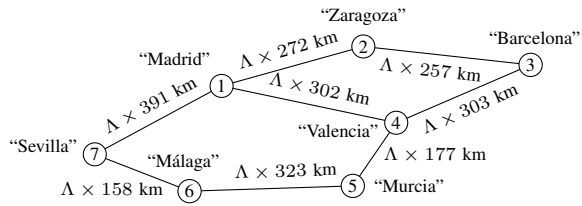

We conducted simulations using the open-source network planning tool
Net2Plan~\cite{pavon-2015-net2plan}, in the offline network design mode,
using an algorithm described below.
We started from an example topology and traffic matrix representing
a ``Fictitious mesh network connecting the seven most populated cities in Spain'',
(Fig.~\ref{fig/topology}),
with 7 nodes and 16 unidirectional links, and 42 traffic demands.
A traffic demand is a request to carry a given amount of traffic
(its ``offered traffic'') between specified origin and destination
node pairs.  The 42 demands in this example offer between 6.95 and 1815~Gbit/s,
for a total offered traffic (summed over all demands) of 10~Tbit/s.
The links average 273~km each; however, since CV-QKD is usually limited
to ranges up to tens of km, we rescaled all the links by a uniform factor
$\Lambda$, using values between 0.01 and 0.25.
For instance, with $\Lambda = 0.01$ (which could represent a campus-scale
network), the longest link (1--7) is 3.9~km
long and can sustain 345~Mbit/s of CV-QKD traffic; this value drops to
88~Mbit/s with $\Lambda = 0.05$ (19.6~km, e.g. city-scale);
and 0 with $\Lambda = 0.1$ (39.1~km), but traffic can still be relayed through other paths.
Similarly, we kept the same traffic matrix for the classical and CV-QKD traffics,
uniformly scaling the total offered traffic to vary the load.
For each value of the parameters, and of the classical and CV-QKD offered traffic,
quantum and classical channels are allocated as described in the following
sections.  The CV-QKD is allocated first, getting priority over the classical
traffic, as we aim to study the impact of a given QKD traffic on the classical
capacity.

\subsection{CV-QKD traffic routing heuristic: opaque}

CV-QKD demands are routed as in an opaque network:
given that CV-QKD is range-limited, we do not attempt any optical bypass,
which would lengthen the paths and reduce capacity.  Instead, we assume
each node is a trusted relay, which also aggregates the traffic from all
CV-QKD demands that need the same link.
Each link's CV-QKD capacity in terms of maximum key rate is calculated
in the absence of Raman noise (because there is no
classical traffic yet).  Then, at each step, the demand with the largest
amount of yet-unallocated traffic is served: the $k$ shortest possible paths
are examined, selecting the one with the largest remaining capacity, and a
route is added carrying as much of the demand's traffic as possible without
increasing the aggregated amount on any link beyond that link's capacity.
This step is then repeated until all CV-QKD demands are served or no more
routes can be added.
Optionally, in some of our simulations, we kept the total CV-QKD traffic
in each link below the link's capacity by a small (e.g. 1~\%) margin;
we will see that
it has a strong impact.

\subsection{Classical traffic RWA heuristic: transparent fixed-grid}

A routing and wavelength assignment (RWA) heuristic is used to
route the classical traffic demands as in a transparent network:
lightpaths are created directly from origin to destination node,
occupying the same WDM channel over all the links in the path
(wavelength continuity), assuming each node is a perfectly flexible
reconfigurable optical add-drop multiplexer (ROADM).  Given the short
ranges, all paths are considered classically feasible and regeneration
is not considered.  Each lightpath occupies 1 WDM slot and carries a
fixed traffic $\mathcal{R}$.  No provision is made to ensure that the
multiple lightpaths required to serve large demands follow the same
route or occupy contiguous WDM slots.

As above, demands are served in decreasing order of yet-unallocated
traffic.  At each step, the $k$ shortest possible paths are examined,
checking that: a WDM slot is available over the whole path; and that
adding a slot to all links of the path would not reduce any link's
CV-QKD capacity below the amount of CV-QKD traffic it carries.
A lightpath carrying 100~Gbit/s is created over the shortest path that
matches both conditions.  This step is repeated until all classical
demands are satisfied or no more lightpaths can be added.

\section{Simulation results}
\label{sec/results}

\begin{figure*}[t!]
  \centering
  \smash{\rlap{\quad(a)}}%
  \includegraphics[width=0.33\linewidth]{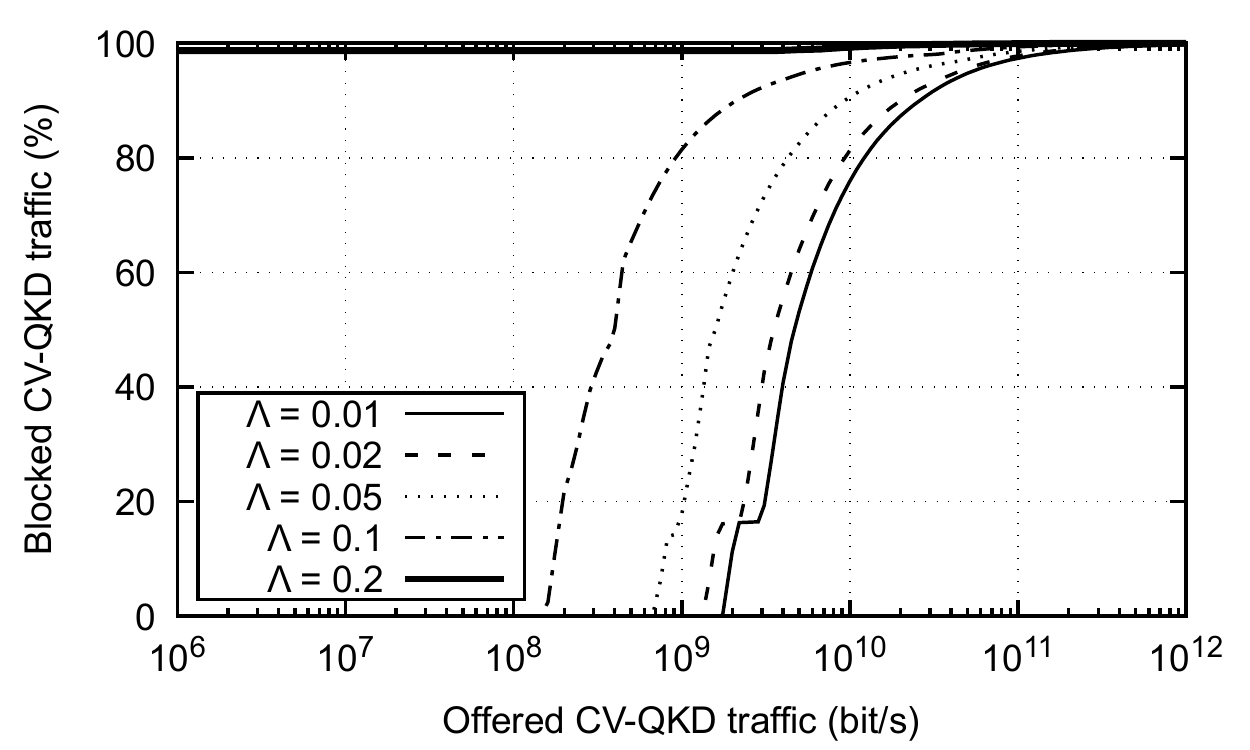}\hfil
  \smash{\rlap{\quad(b)}}%
  \includegraphics[width=0.33\linewidth]{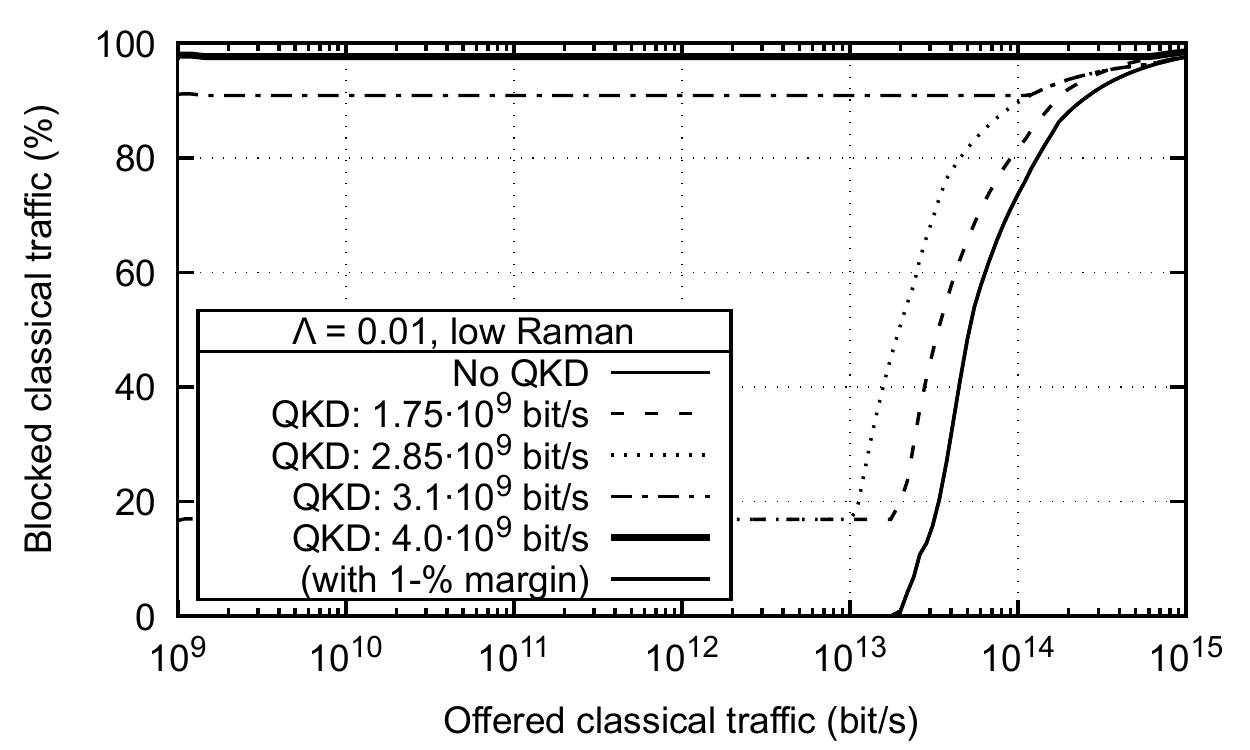}\hfil
  \smash{\rlap{\quad(c)}}%
  \includegraphics[width=0.33\linewidth]{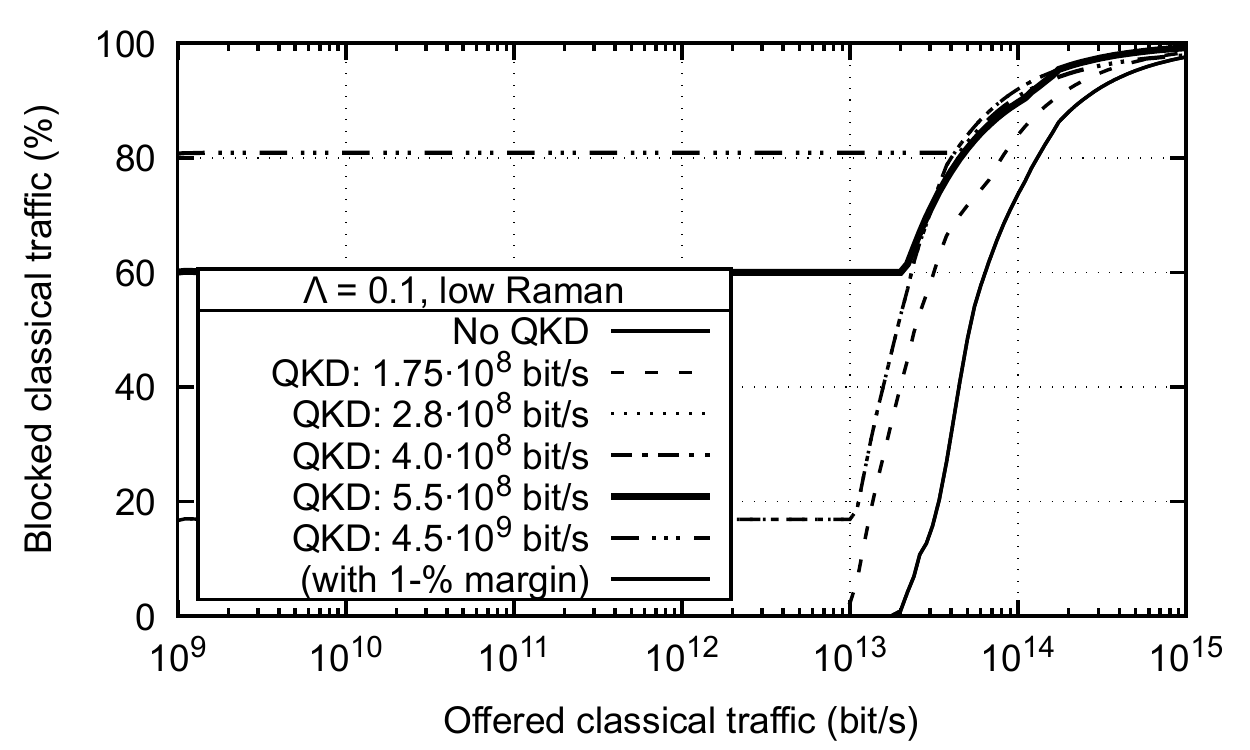}
  \smash{\rlap{\quad(d)}}%
  \includegraphics[width=0.33\linewidth]{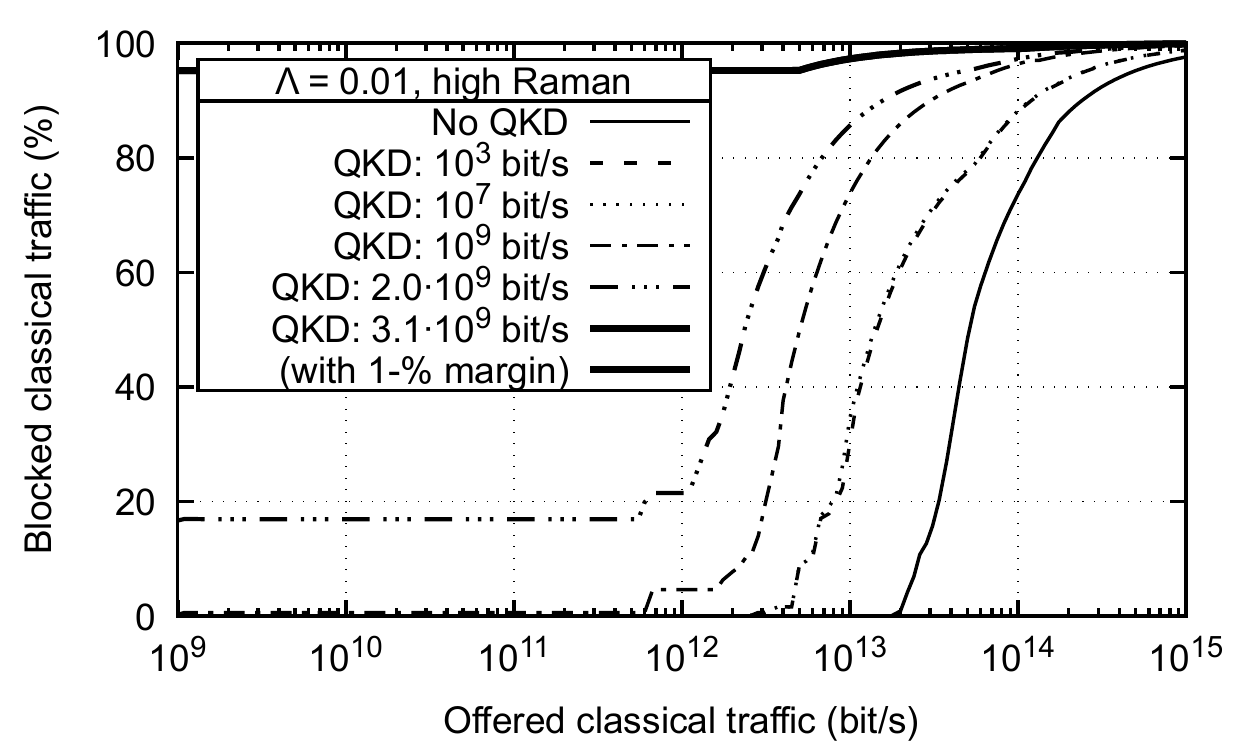}%
  \hfil
  \smash{\rlap{\quad(e)}}%
  \includegraphics[width=0.33\linewidth]{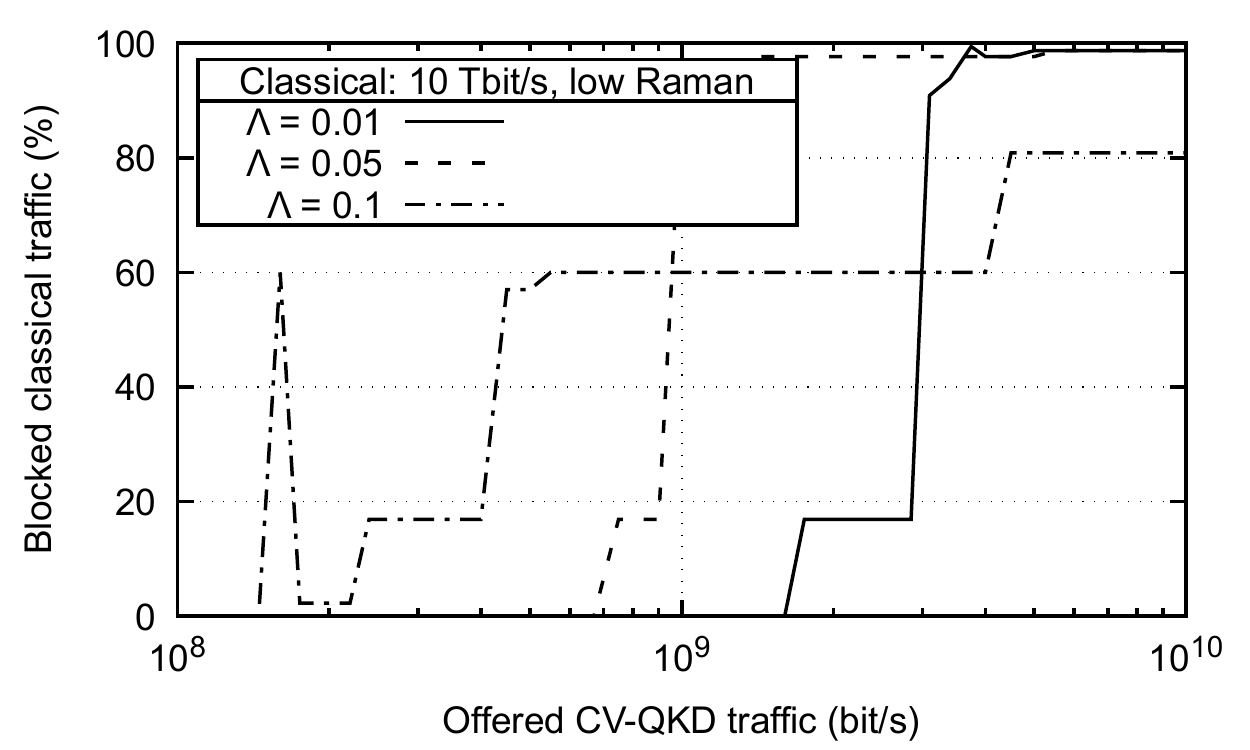}\hfil
  \smash{\rlap{\quad(f)}}%
  \includegraphics[width=0.33\linewidth]{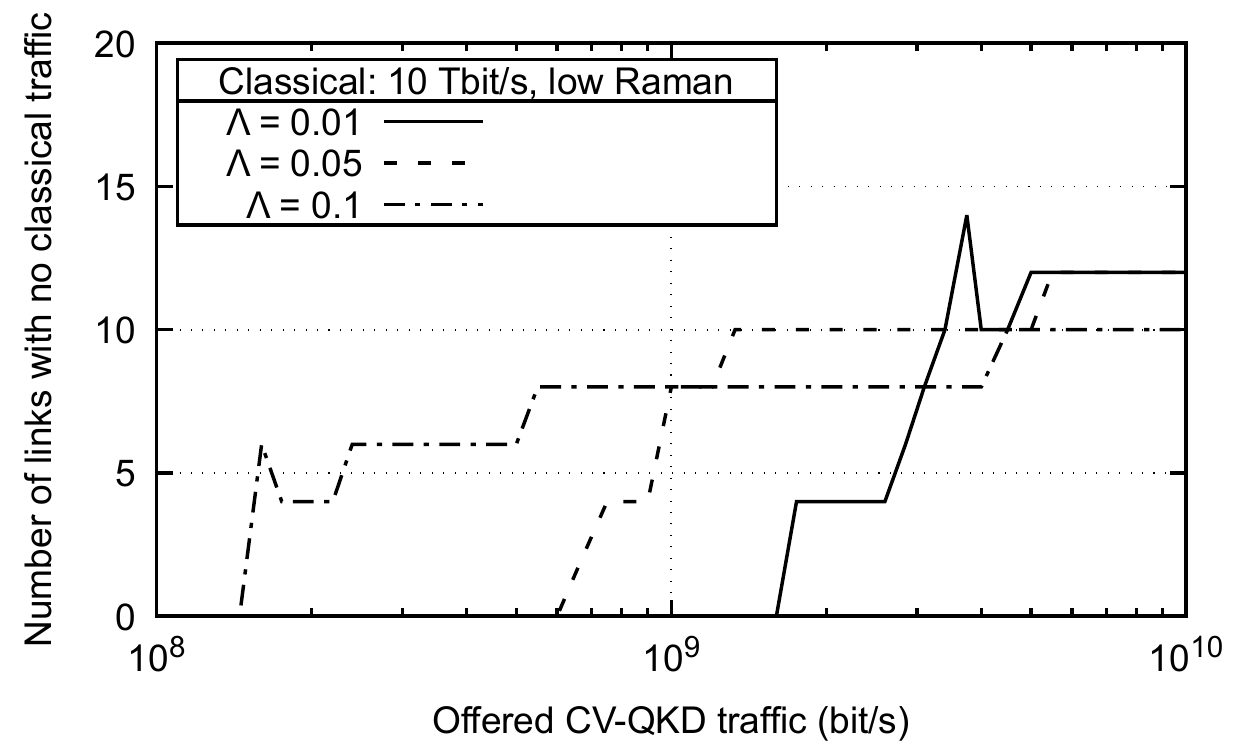}
  \smash{\rlap{\quad(g)}}%
  \includegraphics[width=0.33\linewidth]{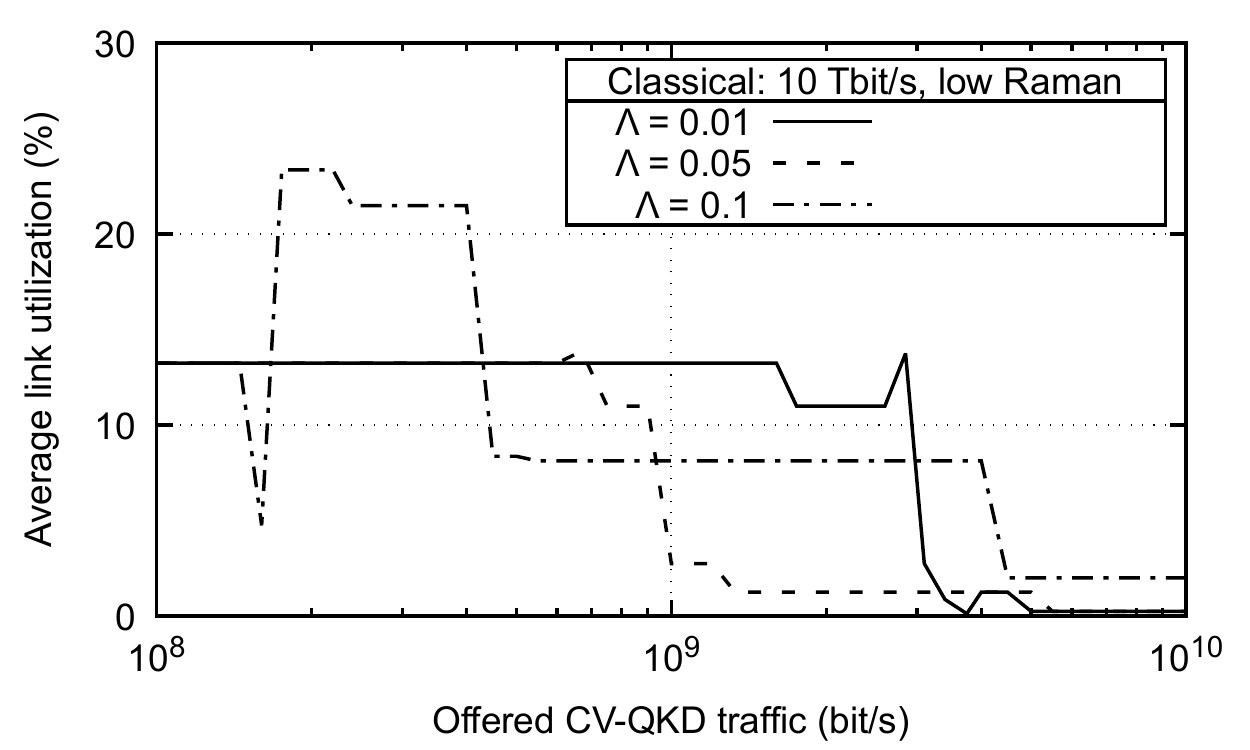}\hfil
  \smash{\rlap{\quad(h)}}%
  \includegraphics[width=0.33\linewidth]{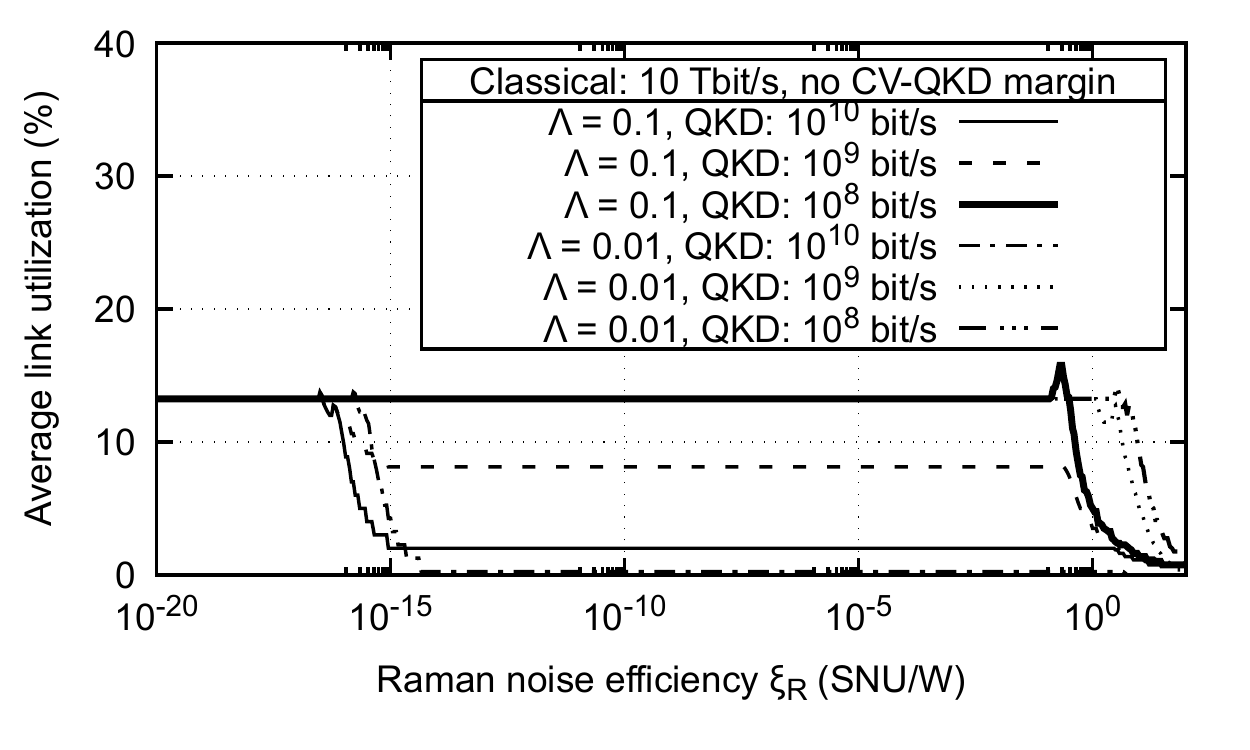}
  \smash{\rlap{\quad(i)}}%
  \includegraphics[width=0.33\linewidth]{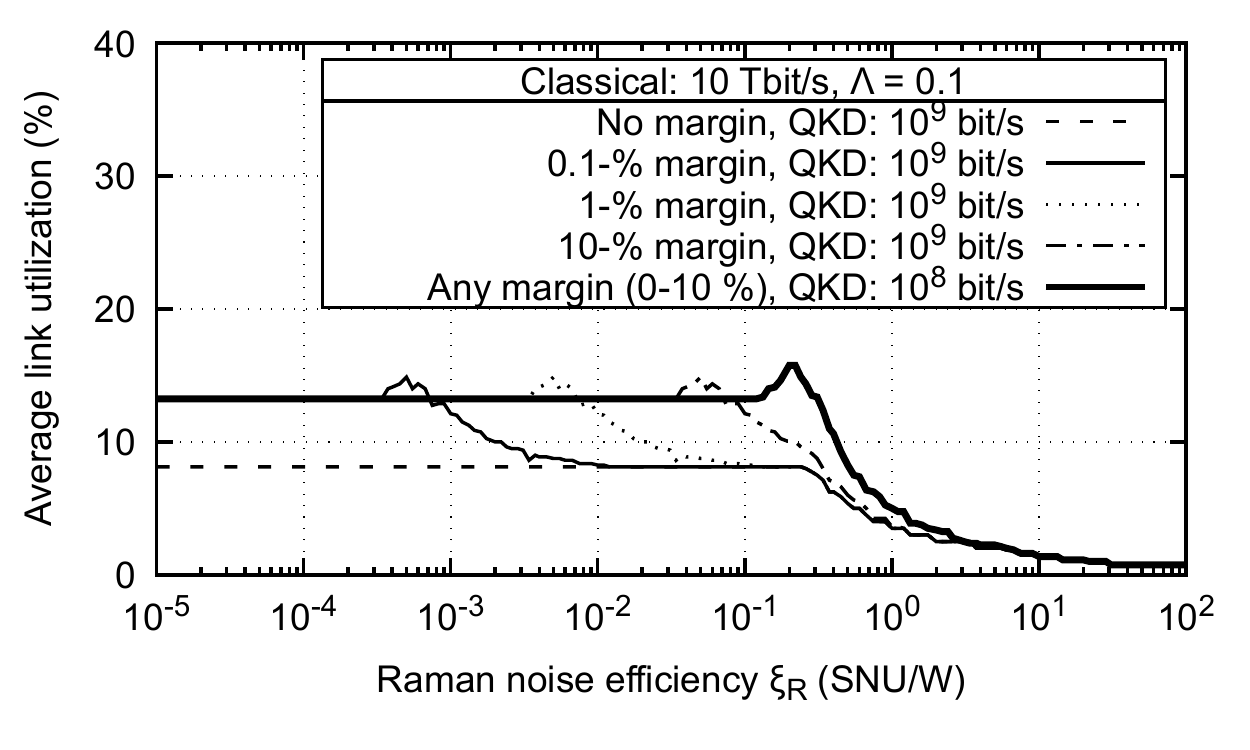}
  \caption{%
    (a)--(e): proportion of traffic blocked (classical or CV-QKD)
    as a function of network load, represented by the total
    offered traffic (classical or CV-QKD).  Multiple values of
    the distance scaling factor $\Lambda$ are tested; also two values
    of the Raman noise efficiency: $\xi_R = 10^{-12}$~SNU/W (``low Raman'')
    or 10~SNU/W (``high Raman'').
    Some plots are repeated with a 1-\% margin in CV-QKD capacity,
    curves overlap with the no-margin curves as indicated by the line style.
    (f)--(g): number of links unused by classical traffic, and average
    link utilization (WDM slots active in the optical fibers), as a
    function of the offered CV-QKD traffic.
    (h)--(i): average link utilization as a function of $\xi_R$,
    without margin in CV-QKD capacity, and with a 0 to 10-\% margin.
    \label{fig/blocking-by-load}
  }
\end{figure*}

\subsection{CV-QKD capacity vs network scale}

Figure~\ref{fig/blocking-by-load}(a) shows the proportion of total
requested CV-QKD traffic that cannot be served without
exceeding capacity on one of the links (thus ``blocked''), as a
function of the total offered CV-QKD traffic.  In our methodology,
this does not depend on the classical traffic or the Raman effect,
since CV-QKD is served first.  It does depend on the distance scaling
factor~$\Lambda$, as links' capacities change with propagation
length.  The network capacity, in terms of total CV-QKD traffic that
can be carried with negligible blocking, is about 0.7--2~Gbit/s for
$\Lambda = 0.01$--$0.05$ (which brings the average link length to a
few km, as in a small-to-large city network), then drops to
100--200~Mbit/s for $\Lambda = 0.1$, above which the distances become
too large for any significant traffic to be carried.

\subsection{Classical capacity vs CV-QKD}

Figure~\ref{fig/blocking-by-load}(b), (c) and (d) show the blocked classical
traffic as a function of the total offered classical traffic, for several
values of the offered CV-QKD traffic, using $\Lambda = 0.01$ or $0.1$,
and two values for $\xi_R$ that will be shown in
Sec.~\ref{sec/raman-coefficient} to be representative:
$10^{-12}$~SNU/W (``low Raman'') and $10$~SNU/W (``high Raman'').
No margin is applied on the CV-QKD capacity, allowing links to carry their
maximum CV-QKD traffic.
The thin solid-line curve gives the classical network
capacity without any CV-QKD, about 20~Tbit/s, independent of $\Lambda$
due to our methodology.  With CV-QKD traffic, below the values shown,
in the low-Raman case,
the curve is identical to the no-QKD case.  Above a certain amount of
CV-QKD traffic, however, a ``floor'' of blocking appears even for very
low amounts of classical traffics.  The floor increases sharply and
discretely with CV-QKD traffic: note that some curves overlap for
different amounts of CV-QKD traffic.  This can also be seen on
Fig.~\ref{fig/blocking-by-load}(d) for the high-Raman case, except
that even with a CV-QKD traffic as low as 1~kbit/s, the classical capacity
is reduced to about 3~Tbit/s.

This discrete behavior also appears on Fig.~\ref{fig/blocking-by-load}(e):
the blocking of classical traffic
increases by steps with the CV-QKD offered traffic, with even the spurious
``spike'' observed at $1.6\cdot 10^8$~bit/s for $\Lambda = 0.1$ reaching
the same level as a later plateau.
On such a small network topology, this seems consistent
with individual links ``dropping out'' of the network, that is, becoming
unavailable to classical traffic.
This is supported by Fig.~\ref{fig/blocking-by-load}(f):
as CV-QKD traffic increases, a number of links stop carrying
classical traffic.
Also, Fig.~\ref{fig/blocking-by-load}(g) shows the average link
utilization (percentage of WDM slots in the fiber that are actually used).
It starts at 13~\%, comparable to the ratio of offered classical traffic
(10~Tbit/s) to theoretical maximum capacity
($N\mathcal{R}\times\text{number of links} = 64$~Tbit/s), which is 16~\%;
then it may increase, as classical traffic is diverted to more circuitous
routes; but eventually it drops as more links become unavailable and
classical demands just can't be met.

The range of CV-QKD traffic where the impact varies most ($10^8$--$10^{10}$~bit/s)
corresponds to where the CV-QKD blocking rate increases, as per
Fig.~\ref{fig/blocking-by-load}(a).  This suggests that the reason
why links become unavailable to classical traffic is that CV-QKD
traffic requires their maximum capacity---at which point adding any
optical power to support classical traffic, even with a low Raman
noise generation, reduces the capacity below the required level,
which is forbidden in our methodology.

This suggests that limiting the CV-QKD traffic below the maximum
by a certain margin might ``unclog'' the links.  We redid some of
the simulations corresponding to Figs.~\ref{fig/blocking-by-load}(b)--(d)
with a 1-\% margin on the links' CV-QKD capacity.  The graphs (partially shown
here) are identical to Fig.~\ref{fig/blocking-by-load}(d) in the high-Raman case,
the margin has no effect; but in the low-Raman case all the curves overlap
the no-QKD curve, that is, the impact of CV-QKD on classical traffic
disappears entirely.

\subsection{Choice of Raman noise efficiency}
\label{sec/raman-coefficient}

In our model, the Raman noise efficiency~$\xi_R$ is a simplistic
assessment of Raman nonlinear scattering.  We will attempt to show
empirically that a more realistic model isn't actually necessary
at this stage of the investigation.

As a starting point, the experimental value from
\cite[Fig.~5]{kumar-2015-coexistence} is about 11~SNU/W for a 25-km
propagation length.  Other references exhibit similar, sometimes lower
values.  Experimental conditions vary, especially the isolation
(spectrally or otherwise) between the quantum and classical signals.
As conditions may improve, we also want to test lower $\xi_R$ values.

Figure~\ref{fig/blocking-by-load}(h) shows the average link utilization
for a classical traffic of 10~Tbit/s, as a function of $\xi_R$, from
$10^{-20}$ to $10^2$~SNU/W.
Three different regimes appear: at very low $\xi_R$ ($\leqslant 10^{-16}$~SNU/W),
link utilization remains high; then a long plateau ($10^{-15}$--$10^{-1}$)
where the impact of CV-QKD on classical traffic remains constant;
and finally the impacts worsens at higher values.
The first regime could be a numerical artifact: at such low values,
floating-point numbers might not have enough precision to show the
drop in CV-QKD key rate due to the Raman noise.  Given the unlikelihood
that Raman noise is so low anyway, we ignored this regime,
and chose to perform
the other simulations with values representative of the other two:
$\xi_R = 10^{-12}$~SNU/W (``low Raman'', on the plateau) and $\xi_R = 10$~SNU/W
(``high Raman'', in the worsening regime).
This seemed sufficient in the no-margin case.
Several simulations were checked with $\xi_R = 10^{-3}$~SNU/W,
and yielded exactly the same results as with $10^{-12}$.

On the other hand, Fig.~\ref{fig/blocking-by-load}(i) shows the same
graph but with several values of the margin on the CV-QKD capacity.
Here link utilization remains at its high starting level up to
a $\xi_R$ that depends on the margin and possibly the CV-QKD traffic,
then starts dropping as before.
This suggests that in some cases (though not e.g. with the
lowest CV-QKD traffic), a margin helps maintain classical capacity
if Raman efficiency is low.

\section{Discussion and questions for future work}
\label{sec/discussion}

The results above indicate that CV-QKD traffic significantly
impacts classical network capacity in some cases, especially
if Raman noise turns out to be high.  However, this study is still
preliminary, opening several potential lines of investigation.

\paragraph*{The specificity of the network topology and traffic matrix}
we only studied one small topology where discrete effects can be seen
when optical links become unavailable for classical traffic.  It would
be interesting to identify which links are affected first; at a guess,
the longest links should be more vulnerable.  Also, in a more complex
topology, especially one more densely-meshed, would the higher number
of alternate paths between nodes compensate the drop in capacity?

\paragraph*{Keeping a CV-QKD capacity margin or using smarter heuristics}
limiting the CV-QKD traffic in each link slightly below the maximum
capacity seems to help, but how much margin is needed remains to be
investigated.  Given that it depends on the Raman efficiency, this
may require a more realistic model as well.
Alternatively, more advanced planning heuristics and
design tools could find better compromises between classical and CV-QKD traffic,
e.g.\ with priority management between the CV-QKD and classical traffic matrices.

\paragraph*{What is the actual Raman noise, if needed}
our linear model may be sufficient to show that an impact exists,
but it does not accurately represent the physical reality of
Raman noise generation.  The question is open of how much accuracy
is actually needed: the value of the required CV-QKD capacity margin
certainly depends on the Raman noise, but is this dependence very
sensitive, or could a rough estimate be satisfactory?
The answer will certainly also depend on the spectral distance
between the CV-QKD signal and the classical WDM channels---which may
in turn determine whether CV-QKD can be carried in wavelengths close
to standard WDM bands or a completely separate spectral band must be
alloted.

\section{Conclusion}
\label{sec/conclusion}

We have shown in network-planning simulations that coexistence
of CV-QKD and classical traffic in the same optical fibers can
in some cases have an adverse impact on the classical network capacity,
due to the need to preserve CV-QKD traffic from excessive Raman noise.
Precautions can be taken, such as limiting the amount of CV-QKD traffic
slightly below each optical link's maximum capacity, but how much margin
is required remains an open question.  We have pointed out this and other
potential lines of investigation.  In any case, it seems that designing
a mixed classical and CV-QKD network will require dedicated planning
heuristics and tools.

\section*{Acknowledgments}

This project has received funding from the European Union's Horizon 2020
research and innovation programme under grant agreement No~820466
(Quantum-Flagship project CiViQ: Continuous-Variable Quantum Communications).
Also, this research was partially supported by Labex DigiCosme
(project ANR-11-LABEX-0045-DIGICOSME), operated by ANR as part of the program
``Investissement d'Avenir'' Idex Paris-Saclay (ANR-11-IDEX-0003-02).

\bibliography{IEEEabrv,ware-qkd-network}

\providecommand{\href}[2]{#2}\begingroup\raggedright\begin{thebibliography}{1}

\bibitem{kumar-2015-coexistence}
R.~Kumar, H.~Qin, and R.~All{\'e}aume, ``Coexistence of continuous variable
  {QKD} with intense {DWDM} classical channels,'' {\em New Journal of Physics}
  {\bfseries 17} no.~4, (2015) 043027,
  \href{http://arxiv.org/abs/1412.1403}{{\ttfamily arXiv:1412.1403
  [quant-ph]}}.

\bibitem{eriksson-2019-wdm-cvqkd}
T.~A. Eriksson, T.~Hirano, B.~J. Puttnam, G.~Rademacher, R.~S. Lu{\'\i}s,
  M.~Fujiwara, R.~Namiki, Y.~Awaji, M.~Takeoka, N.~Wada, {\em et~al.},
  ``Wavelength division multiplexing of continuous variable quantum key
  distribution and 18.3 {T}bit/s data channels,'' {\em Communications Physics}
  {\bfseries 2} no.~9, (2019) 1--8.

\bibitem{aymeric-OFC-2022-symbiotic}
R.~Aymeric, Y.~Jaouën, C.~Ware, and R.~Alléaume, ``Symbiotic joint operation
  of quantum and classical coherent communications,'' in {\em Optical Fiber
  Communication Conference}, no.~W2A.37.
\newblock Mar., 2022.
\newblock \href{http://arxiv.org/abs/2202.06942}{{\ttfamily arXiv:2202.06942
  [quant-ph]}}.
\newblock Poster.

\bibitem{laudenbach-cvqkd-theory-2018}
F.~Laudenbach, C.~Pacher, C.-H.~F. Fung, A.~Poppe, M.~Peev, B.~Schrenk,
  M.~Hentschel, P.~Walther, and H.~H{\"u}bel, ``Continuous-variable quantum key
  distribution with gaussian modulation—the theory of practical
  implementations,'' {\em Advanced Quantum Technologies} (2018) 1800011,
  \href{http://arxiv.org/abs/1703.09278}{{\ttfamily arXiv:1703.09278
  [quant-ph]}}.

\bibitem{azuma-2021-quantum-network-design}
K.~Azuma, S.~B{\"a}uml, T.~Coopmans, D.~Elkouss, and B.~Li, ``Tools for quantum
  network design,'' {\em AVS Quantum Science} {\bfseries 3} no.~1, (2021)
  014101, \href{http://arxiv.org/abs/2012.06764}{{\ttfamily arXiv:2012.06764
  [quant-ph]}}.

\bibitem{pavon-2015-net2plan}
P.~Pavon-Marino and J.-L. Izquierdo-Zaragoza, ``Net2plan: an open source
  network planning tool for bridging the gap between academia and industry,''
  {\em {IEEE} Netw.} {\bfseries 29} no.~5, (2015) 90--96.

\end{thebibliography}\endgroup
\bibliographystyle{utphys}


\end{document}